\documentclass{svproc} 

\usepackage[top=5.2cm, bottom=5.2cm, left=4.4cm, right=4.4cm]{geometry}

\usepackage{graphicx} 
\graphicspath{ {./images/} }

\usepackage{tabularx}

\title{Taxonomy of migration scenarios for Qiskit refactoring using LLMs}

\author{José Manuel Suárez \and Luís Mariano Bibbó \and Joaquín Bogado \and Alejandro Fernandez}
\institute{Universidad Nacional de La Plata, Facultad de Informática 
\newline Laboratorio de Investigación y Formación en Informática Avanzada (LIFIA), Argentina}

\date{February 2025}

\begin{document}

\maketitle
 
\begin{abstract}
As quantum computing advances, quantum programming libraries' heterogeneity and steady evolution create new challenges for software developers. Frequent updates in software libraries break working code that needs to be refactored, thus adding complexity to an already complex landscape. These refactoring challenges are, in many cases, fundamentally different from those known in classical software engineering due to the nature of quantum computing software. This study addresses these challenges by developing a taxonomy of quantum circuit's refactoring problems, providing a structured framework to analyze and compare different refactoring approaches. Large Language Models (LLMs) have proven valuable tools for classic software development, yet their value in quantum software engineering remains unexplored. This study uses LLMs to categorize refactoring needs in migration scenarios between different Qiskit versions. Qiskit documentation and release notes were scrutinized to create an initial taxonomy of refactoring required for migrating between Qiskit releases. Two taxonomies were produced: one by expert developers and one by an LLM. These taxonomies were compared, analyzing differences and similarities, and were integrated into a unified taxonomy that reflects the findings of both methods. By systematically categorizing refactoring challenges in Qiskit, the unified taxonomy is a foundation for future research on AI-assisted migration while enabling a more rigorous evaluation of automated refactoring techniques. Additionally, this work contributes to quantum software engineering (QSE) by enhancing software development workflows, improving language compatibility, and promoting best practices in quantum programming. This research marks the first step in a broader effort to assess various refactoring strategies, ultimately guiding the development of AI-powered tools to support quantum software engineers.
\end{abstract}

\vspace{10pt} 
\textbf{Keywords:} Quantum Computing (QC), Quantum Software Engineering (QSE), \\ Large Language Models (LLMs), Generative AI, Qiskit, Migration Code.

\newpage

\section{Introduction}
The recent advancements in quantum computing (QC) \cite{preskill_quantum_2018} underscore the necessity for robust methodologies within Quantum Software Engineering (QSE)\\ \cite{jimenez-navajas_code_2025}. Consequently, well-established topics in classical Software Engineering (SE)—such as code generation \cite{jin_towards_2025} \cite{asif_pennylang_2025}\\ \cite{dupuis_qiskit_2024}, refactoring \cite{zhao_refactoring_2023} \cite{tsantalis_refactoringminer_2022}, metric definition\\ \cite{nation_benchmarking_2025} \cite{quetschlich_mqt_2023} \cite{zhao_size_2021}, algorithmic strategy design \cite{xu_logic_2018}, development stacks \cite{guo_isq_2023}, debugging and testing strategies \cite{ramalho_testing_2024} \cite{ali_assessing_2021} \cite{li_proq_2020}, and languages and compilers\\ \cite{amy_staq_2020} \cite{sivarajah_tketrangle_2021}—must now be re-examined within the QSE framework, demanding more sophisticated approaches.

Specifically, the complexity of processes such as code migration and the analysis of refactoring scenarios, despite being widely studied topics within the framework of SE, require a novel approach. Such is the case with the challenges present in highly dynamic development ecosystems like Qiskit, whose evolution imposes a release schedule that often affects the functionality of algorithms. In addition, its proximity to QC as an interdisciplinary development field adds the difficulty that not all those involved are experts in SE areas.

At the same time, the rise of generative AIs \cite{weisz_design_2024} \cite{bengesi_advancements_2023} and large language models (LLMs) in particular~\cite{openai_gpt-4_2024} \\ \cite{team_gemini_2024}
\cite{roziere_code_2024} 
presents an unprecedented opportunity for automation and optimization of code migration and refactoring processes. However, their full integration into the QSE development cycle \cite{murillo_quantum_2025} is still in early stages, with unresolved challenges, especially in terms of correctness and reliability.

The relevance of our work revolves around the development of a resolutive and at the same time replicable hybrid methodology that integrates SE strategies and the use of LLMs, based on the analysis of migration and refactoring scenarios in Qiskit, in order to assess their feasibility and summarize their key factors and implications. Complementarily, this study aims to be the initial step toward the development of tools that enhance workflow agility and reduce the technological gap through the early adoption of more recent and effective functionalities in the QSE domain.

The evolution of QC has been directly linked to the accelerated maturity of quantum development environments such as Qiskit \footnote{IBM Qiskit official site - \url{https://www.ibm.com/quantum/qiskit}} \cite{javadi-abhari_quantum_2024}, \\ Cirq \cite{isakov_simulations_2021}, Quipper \cite{green_quipper_2013}, Q\# \cite{svore_q_2018}, PennyLane\footnote{Xanadu Pennylane - \url{https://docs.pennylane.ai/en/stable/}}, among others; they have played a fundamental role regarding accessibility and quality in the development, simulation, execution, and deployment process of quantum algorithms, increasing their versatility and successfully integrating with external libraries and different quantum technologies. In this work, we will particularly focus on the Qiskit ecosystem, launched in 2017 by IBM, an open-source environment compatible with Python, backed by a broad community, which supports the simulation and real execution of quantum algorithms, integration with external environments and extensions (Qiskit extensions), platforms and hardware architectures, support for Quantum Machine Learning (QML) \cite{Biamonte_2017} and hybrid execution, among other features, making it one of the most relevant development ecosystems today.
\footnote{IBM Quantum Documentation (\url{https://docs.quantum.ibm.com/})}

The structure of this work proceeds as follows: in Section~\ref{sec:relatedwork} addresses related work, allowing us to more precisely establish the current state of research; Section~\ref{sec:methodology} then focuses on a detailed description of the experimental methodology used; in Section~\ref{sec:results} we present the preliminary results; Section~\ref{sec:discussion} outlines the main discussions in the field, concluding in Section~\ref{sec:futurework} with some conclusions and future work and research guidelines that should follow in the short and medium term.

\section{Related work}
\label{sec:relatedwork}
We can subdivide the works analyzed in this section according to the use of automatic generation tools and based on their relation to QC, in addition to a handful of works that combine both subgroups:

At the intersection of automatic code migration using LLMs, we can mention recent significant efforts \cite{almeida_automatic_2024} where the work related to prompts engineering is significant. Even if some of the test cases involve some subjectivity that could introduce biases and unreliable conclusions, they allow pondering the implication of prompts in the metrics obtained; on the other hand \cite{sahoo_systematic_2024} provides a very complete review on prompt engineering and the most relevant associated works, considering query techniques, contextual refinement and precision on the associated metrics, as well as prompt iteration and rephrasing \cite{deng_rephrase_2024}. Other works aligned with our research promote the generation of surpassing LLMs \cite{cordeiro_empirical_2024} focused on the Java language, although our case study is more complex due to its recent expansion, high dynamism and limited code sources, in addition to difficulties inherent to the quantum paradigm and its distinctive properties.

At the intersection of QC and migration, we consider the work of \cite{zhao_refactoring_2023}, focused on the challenges of refactoring in quantum programs on Q\#, although limited to maintainability and efficiency criteria and not so closely related to the migration process between versions; likewise, it proposes a catalog based on the analysis of algorithms, similar to our taxonomy, and just as their proposal of \textit{QSharp Refactoring Tool}, our work aims at the construction of a hybrid automated migration tool.

We also consider works related to Qiskit \cite{dupuis_qiskit_2024} because of its closeness to the framework we are studying, in addition to some few works that relate LLMs strategies together with quantum code migration (\cite{asif_pennylang_2025}) where RAG \cite{lewis2021retrievalaugmentedgenerationknowledgeintensivenlp} is used for a code generation wizard over PennyLane and evaluation metrics. While these works address LLMs code generation performance and evaluation, do not address the specifics of version migration and the refactoring issues that arose from them. 
 
\section{Methodology}
\label{sec:methodology}
\subsection{General description}

\begin{figure}[h!]
    \centering    
    \includegraphics[width=0.5\textwidth]{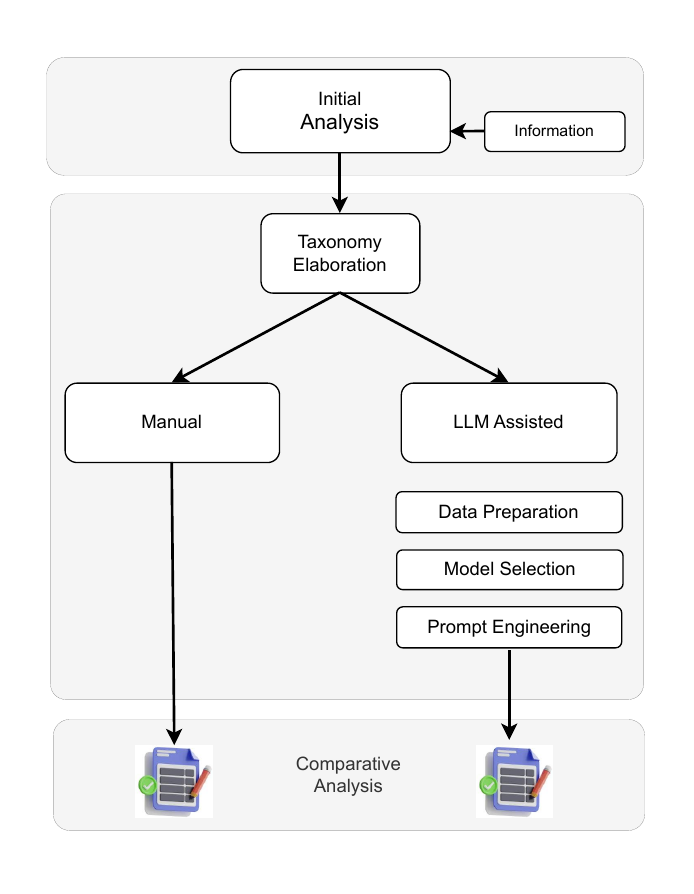}
    \caption{Experimental flow diagram}
    \label{fig:flujo_experimental}
\end{figure}

The methodological structure of our experiment is divided into three fundamental stages: the first consists of the collection of prior and necessary supporting information to establish the experiment, then a forked into a manual and automated LLM assisted stages for the elaboration of a taxonomy of migration scenarios on Qiskit. Each taxonomy section is Qiskit version specific. The third stage relates to the comparison of the taxonomies obtained in the previous stage. See Figure~\ref{fig:flujo_experimental}.


\subsection{Manual elaboration of taxonomy of migration scenarios}

For the elaboration of the taxonomies we used the official sources of IBM Qiskit exclusively. We consider this documentation authoritative, accurate and detailed. This source of documentation maintains high precision in terms of API changes, obsolescence and deprecation warning flows, alternatives, recommendations, accompanied by concrete examples and best practice guidelines.

In this regard, we consider two main official sources:
\begin{itemize}
    \setlength{\itemindent}{1cm}
    \item Qiskit release notes. 
    \footnote{Qiskit SDK release notes (\url{https://docs.quantum.ibm.com/api/qiskit/release-notes})}
    \item Qiskit changelog. 
    \footnote{Qiskit Changelog (\url{https://github.com/qiskit/qiskit/releases})}
\end{itemize}

We consider secondary sources in the manual stage to allow for disambiguation or additional explanation in cases we saw fit. These secondary sources are:
\begin{itemize}
    \setlength{\itemindent}{1cm}
    \item Qiskit Leatest updates.
    \footnote{Qiskit Leatest updates (\url{https://docs.quantum.ibm.com/guides/latest-updates})}
    \item Qiskit Documentation GitHub.
    \footnote{Qiskit Documentation GitHub (\url{https://github.com/Qiskit/documentation/tree/main})}
    \item Qiskit Migration guides.
    \footnote{Qiskit Migration guides (\url{(https://docs.quantum.ibm.com/migration-guides})}
    \item Qiskit release summary.
    \footnote{Qiskit release summary (\url{https://www.ibm.com/quantum/blog/qiskit-1-0-release-summary})}
    \item Youtube Qiskit chanel.
    \footnote{Youtube Qiskit chanel (\url{https://www.youtube.com/qiski})}
\end{itemize}

Regarding the information supported by the taxonomy, it seemed important to classify each migration scenario based on its functionality or objective. Many times these categories have arisen from the documentation itself and others based on well-known areas of refactoring within the SE field \cite{zhao_refactoring_2023}, see Table \ref{tab:tabla_categorizacion_escenarios}. 

\begin{table}[htbp]
\centering
\caption{Manual taxonomy - Categorization of scenarios}
\renewcommand{\arraystretch}{1.2} 
\setlength{\tabcolsep}{5pt} 
\label{tab:tabla_categorizacion_escenarios}
\resizebox{0.8\textwidth}{!}{ 
    \begin{tabularx}{\linewidth}{|>{\hspace{2pt}}p{2.5cm}|X|p{3.5cm}<{\hspace{3pt}}|}
    \hline
    \textbf{Category} & \textbf{Possible values} & \textbf{Description} \\ \hline
    Structure & Reorganization, Package, Module, Inheritance, etc. & Modular ecosystem reorganization, functionality reassignment, new modules or deprecations, etc. \\ \hline
    Language & Python, Rust, C++ & Scenario where the underlying language has changed, usually for performance reasons. \\ \hline
    Implication & Primitive, Gate, Algorithm, Transpilation, Organization, Parameterization, Class, Method & Degree of relationship with the artifact affected. \\ \hline
    Module & Q-Aer, Q-Aqua, Q-Terra, Q-Ignis, Q-IBMQuantum, Q-Experiments, Q-Chemistry, Q-Optimization, Q-Finance, Q-ML, Q-Nature, etc. & Qiskit ecosystem module affected. \\ \hline
    Type & Increased functionality, upgrade, deprecation, bug fixes & Categorization from release notes. \\ \hline
    SE Refactoring & Usability, extensibility, readability, modularity, security, flexibility, etc. & Classic software engineering affectations. \\ \hline
    \end{tabularx}
}
\end{table}

Another dimension that the taxonomy addresses is the degree of difficulty that each migration scenario entails. Although this concept is relatively difficult to evaluate due to its subjectivity, we made an effort to apply it as accurately as possible. We consider lines of code required, time needed to implement the refactoring, atomicity or chaining of changes, and complexity related to QSE specifics. This classification is related to the impact of the change from the perspective of a developer i.e.: with role migrator, associated to the question: \textit{"how complex is it to make progress on the migration scenario?"}, see Table \ref{tab:tabla_categorizacion_grados_dificultad}.

\begin{table}[htbp]
\centering
\caption{Manual taxonomy - Categorization of degrees of difficulty}
\renewcommand{\arraystretch}{1.2} 
\setlength{\tabcolsep}{5pt} 
\label{tab:tabla_categorizacion_grados_dificultad}
\resizebox{0.8\textwidth}{!}{ 
    \begin{tabularx}{\linewidth}{|>{\hspace{2pt}}p{2.5cm}|X<{\hspace{3pt}}|} 
    \hline
    \textbf{Category} & \textbf{Description} \\ \hline
    Minimal & No migration difficulties. \newline e.g.: internal performance optimization. \\ \hline
    Low & Can be migrated without significant effort, little time and code consumed. \newline e.g.: new method parameterization. \\ \hline
    Moderate & Involves considerable effort. \newline e.g.: change in simulation flow or job execution. \\ \hline
    High & Implementing the change is of significant difficulty, requiring chained changes on different sections of the code, considerable time and test case execution. \newline e.g.: new module with replacement functions. \\ \hline
    \end{tabularx}
}
\end{table}

We considered relevant to evaluate the degree of relationship that each scenario maintains with the field of classical and quantum software engineering. This allows us to weight each version in relation to its QSE impact and lays the foundations for a tool that considers the migration impact and QSE relevance. See Table \ref{tab:tabla_impacto_qse}.

\begin{table}[htbp]
\centering
\caption{Manual taxonomy - Degree of impact on SE and/or QSE}
\renewcommand{\arraystretch}{1.2} 
\label{fig:table_1}
\setlength{\tabcolsep}{6pt} 
\label{tab:tabla_impacto_qse}
\resizebox{0.8\textwidth}{!}{ 
\begin{tabularx}{\linewidth}{|>{\hspace{2pt}}p{2.5cm}|X<{\hspace{3pt}}|} 
\hline
\textbf{Category} & \textbf{Description} \\ \hline
Classic Software Engineering (SE) & The scenario affects aspects linked to classic software engineering. \newline e.g.: modularity, extensibility, etc. \\ \hline
Quantum Software Engineering (QSE) & The scenario has implications on QC related issues. \newline e.g.: transpilation, gates, primitives, simulation, integration with external services, etc. \\ \hline
\end{tabularx}
}
\end{table}

To conclude, the general outline of the manual taxonomy includes the following columns. See  Table \ref{tab:tabla_tax_manual}.:
\begin{table}[htbp]
\centering
\caption{Manual taxonomy - classification dimensions}
\renewcommand{\arraystretch}{1.2} 
\setlength{\tabcolsep}{6pt} 
\label{tab:tabla_tax_manual}
\resizebox{0.8\textwidth}{!}{ 
\begin{tabularx}{\linewidth}{|>{\hspace{2pt}}p{2.5cm}|X<{\hspace{3pt}}|} 
\hline
\textbf{Column} & \textbf{Description} \\ \hline
Category & Type of each migratory scenario. \\ \hline
Migration Flow & Source and Target of the version upgrade associated with the scenario. \\ \hline
Summary & Brief description of the scenario. \\ \hline
Artifacts & Summary of keywords of the artifacts involved. \\ \hline
Example code in source version & Python example code in the source version of the flow or previous versions. \\ \hline
Example code in target version & Python example code in the target version of the flow or later. \\ \hline
Degree of Difficulty & Complexity weighting involved in the migration scenario. \\ \hline
Degree of impact in SE/QSE & Relationship of the migrating scenario to aspects of classical software engineering and/or associated with quantum software engineering. \\ \hline
References & Links to the authoritative sources of the scenario, either primary or secondary. \\ \hline
\end{tabularx}
}
\end{table}
We considered the version updates 0.46.0 and 1.0.0 exhaustively. Most of the changes between minor versions do not contain rich refactoring scenarios compared to the mentioned update. We exclude \textit{bug-fixes}, \textit{patch} and \textit{prelude} releases due to these changes do not impact on any refactoring scenario. The update from 0.46.0 to 1.0.0 represents a major version bump\footnote{At the time of writing this article March 2025 where v2.0.0 is in development.}.
We analyzed 13 version updates, with special attention to the releases 0.46.0 and 1.0.0.

\subsection{Automatic elaboration of taxonomy of migration scenarios}

For the automatic taxonomy building stage, we decided to initially perform local tests by downloading models Google gemma and DeepSeek using the LMStudio application \footnote{\url{https://lmstudio.ai/}}:
\begin{itemize}

    \item \textbf{Google gemma-3-27b-it-GGUF} \footnote{\url{https://docs.api.nvidia.com/nim/reference/google-gemma-3-27b-it}}, released in March 2025. In principle, with high accuracy and efficient management of hardware resources that enable its execution on a wide range of devices, the possibility of extended token windows that enable the processing of large amounts of text in a single prompt (128k), allowing more flexible and extensible work approaches. In short, it is a tool that is currently in the state of the art, in terms of lightweight open source models.
    
    \item \textbf{DeepSeek-R1-Distill-Qwen-32b-GGUF} \footnote{\url{https://huggingface.co/bartowski/DeepSeek-R1-Distill-Qwen-32B-GGUF}}, a distilled Qween-32B model fine tuned with the outputs of DeepSeek-r1. This model is a 32 billion parameter model, available under the Apache 2.0 license. It support 32k token context. 

\end{itemize}
We used the quantized to 8 bits version of the models. We loaded the models in a computer with a single AMD Ryzen 9 7950X3D 16-Core Processor, with 128 GB of RAM, and 2x NVIDIA RTX 3090 24 GB VRAM. With GPU offloading the models achieve between 10 to 20 tokens per second. 

This initial phase allowed us to improve the prompts without the need monetary expenses, while establishing a baseline prompts for the commercial models. We tested four of the latest models available commercially: OpenAI's \textbf{gpt-4o} \footnote{\url{https://platform.openai.com/docs/models/gpt-4o}}, OpenAI's \textbf{gpt-4o-mini} \footnote{\url{https://platform.openai.com/docs/models/gpt-4o-mini}}, \textbf{DeepSeek-V3} \footnote{\url{https://api-docs.deepseek.com/news/news250325}} and \textbf{DeepSeek-R1} \footnote{\url{https://api-docs.deepseek.com/news/news250120}}. The models where instructed to generate a taxonomy using markdown with the same structure described in the previous section, over the same Qiskit versions. We used few shot learning to enforce the format of the desired output. 

As previously mentioned, among the main axes of our experiment are the replicability, accessibility and improvement of the experiment through the GitHub repository and the development of a script that allows the parameterization of the test language models, internal parametric configurations, as well as data traceability through the download of collected data and stages of verification of the complete observability of source data by the model, which are the same as those considered by the manual taxonomy generation agent.
In order to guaranty the replicability, we wrote a series of Python scripts that allows us to reproduce the experiments. The scripts tasks are divided into "documentation extraction", "documentation verification" and, "model access". The document extraction retrieves the documentation from the primary source and is parametrized by the version number and language \footnote{The Qiskit documentation is localized automatically based on browser preferences of the user. We used the Spanish and English versions.}. The verification script checks the correctness of the retrieval of documentation based on a manual retrieval example. The model access script allows for model selection programmatically, while allows experiments automation using the same prompts to test different LLMs. 

The prompt structure contains a system part describing the task, that is to generate the taxonomy in an specific markdown format with few shot example, defines the column names and expected content, and also include guidelines for the output, i.e.: to restrict scenario grouping, discourage line braking in row lines, avoid replicated scenarios, etc.  The user prompt includes a one-shot example of a row and the information retrieved by the document extraction script for a specific Qiskit version.



A Python script was developed to trigger the stages described above. This script also logs all the information returned by the different models and group it by the target version. We ensure that the information accessed by the LLMs is the same to that used to generate the manual taxonomy.

The scripts and examples are accessible in a GitHub repository, to ensure experimental replicability~\footnote{GitHub Repository(\url{https://github.com/jwackito/qiskit\_llm\_experiment})}.

\section{Preliminary results}

Initial tests didn't produce satisfactory results. Only by refining the prompt structure and wording the models started to return consistent and well formatted output. For example the first results return very few rows, with very few details and incomplete examples and empty cells or duplicate rows. However, when a refined prompt were used, the number of rows tend to match or exceed the manual taxonomy, as well as the level of details and examples without introducing the manual constrains, i.e: ask for a number of rows. 

There was a clear quality improve produced by the introduction of the text of the documentation inside the user prompt. However this methodology is constrained by the context length of the model measured in tokens. This is not a limitation given that the context for modern LLMs ranges from 32K to 128K tokens, while the most extensive Qiskit documentation is about 20K tokens. See Fig. \ref{fig:qiskit_versions}. The mentioned context length limitation can be walk around by the uses techniques such as RAG or context summarization.


\label{sec:results}
\begin{figure}[h]
    \centering    
    \includegraphics[width=0.9\textwidth]{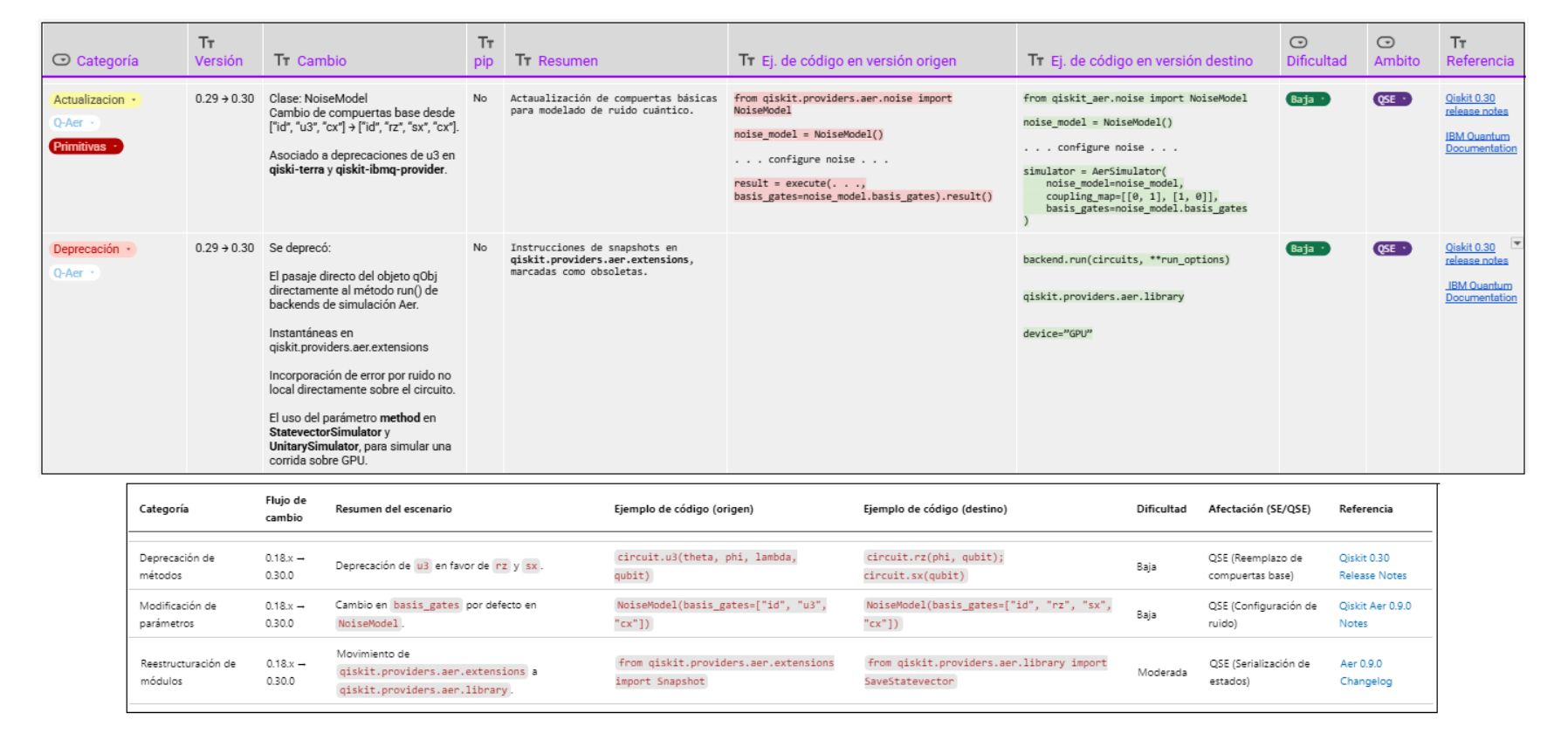}
    \caption{Examples of manual vs. automatic taxonomy scenarios}
    \label{fig:examples}
\end{figure}

In terms of correctness and coherence of scenarios, there is a high correspondence between the manual taxonomy and the output from the commercial models. Regarding the classification of SE vs QSE scenarios, we determined that more complex and information rich prompts allow the models to discern between them with greater precision, providing  appropriate descriptions, with a high degree of alignment with the manual taxonomy. When extracted documentation is included into the prompt, the models behave the best, even surpassing the manual taxonomy in paradigmatic examples. We also noticed a slight improvement in the precision, completeness of examples, and explanation when we used English language prompts. See Fig. \ref{fig:examples}

We observed a a trend in the experiment that goes as follows: \textbf{Initial Phase}, where the scenarios obtained by the manual taxonomy represent a superset of those obtained automatically and the agreement between taxonomies is limited. This is mainly due to rudimentary prompts and limited information included. To only information the model has about the particular problem was gained during training. \textbf{Intermediate Phase}, where prompts include information of the official documentation. In this phase, the intersection between scenarios increase. The LLMs are able to detect scenarios not present in the manual taxonomy or provide a better version of them with improved categorization, richer examples, while still detecting scenarios described by the manual taxonomy. However, this is not the general case yet and some times the models overlooked some important scenarios.

%
%

\begin{figure}[h!]
    \centering    
    \includegraphics[width=0.8\textwidth]{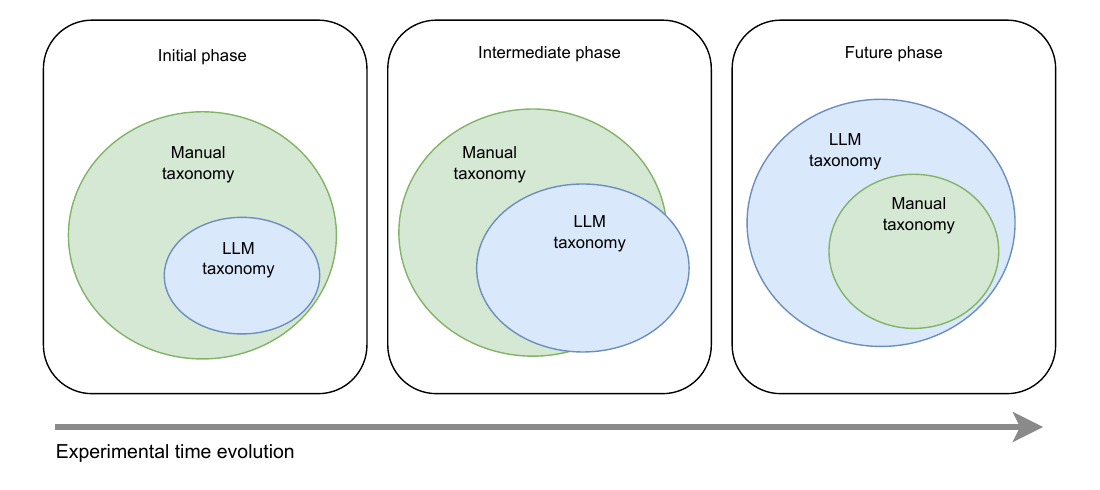}
    \caption{Flow chart of experimental steps}
    \label{fig:flujo_etapas}
\end{figure}

Additionally, we believe the trend flows to an hypothetical \textbf{Future Phase}, as shown in the diagram, see Figure~\ref{fig:flujo_etapas}. In this phase, the LLM models will be able to retrieve all the scenarios of the manual taxonomy while provide information about other scenarios not present in the manual taxonomy.

From what has been previously stated, we can infer that the use of LLMs is plausible and suitable for automatic taxonomy generation of QSE migration scenarios, substantially reducing the required time. This is more clearly evidenced in versions where the documentation is extensive, includes a greater number of disruptive API changes or substantial improvements associated with QSE (usually in major version updates).

\section{Discussion}
\label{sec:discussion}
The creation of a manual taxonomy of migration scenarios is a complex process of review and analysis that requires considerable effort and time. It often requires cross-referencing different authoritative sources, code review, and interaction with other members of the development team. The quality of the final result is directly related to the level and experience of the professional in SE. The use of LLMs to aid in the elaboration of this kind of taxonomies can save time of the SE expert making the process faster and less prone to errors.

\begin{figure}[h!]
    \centering    
    \includegraphics[width=0.9\textwidth]{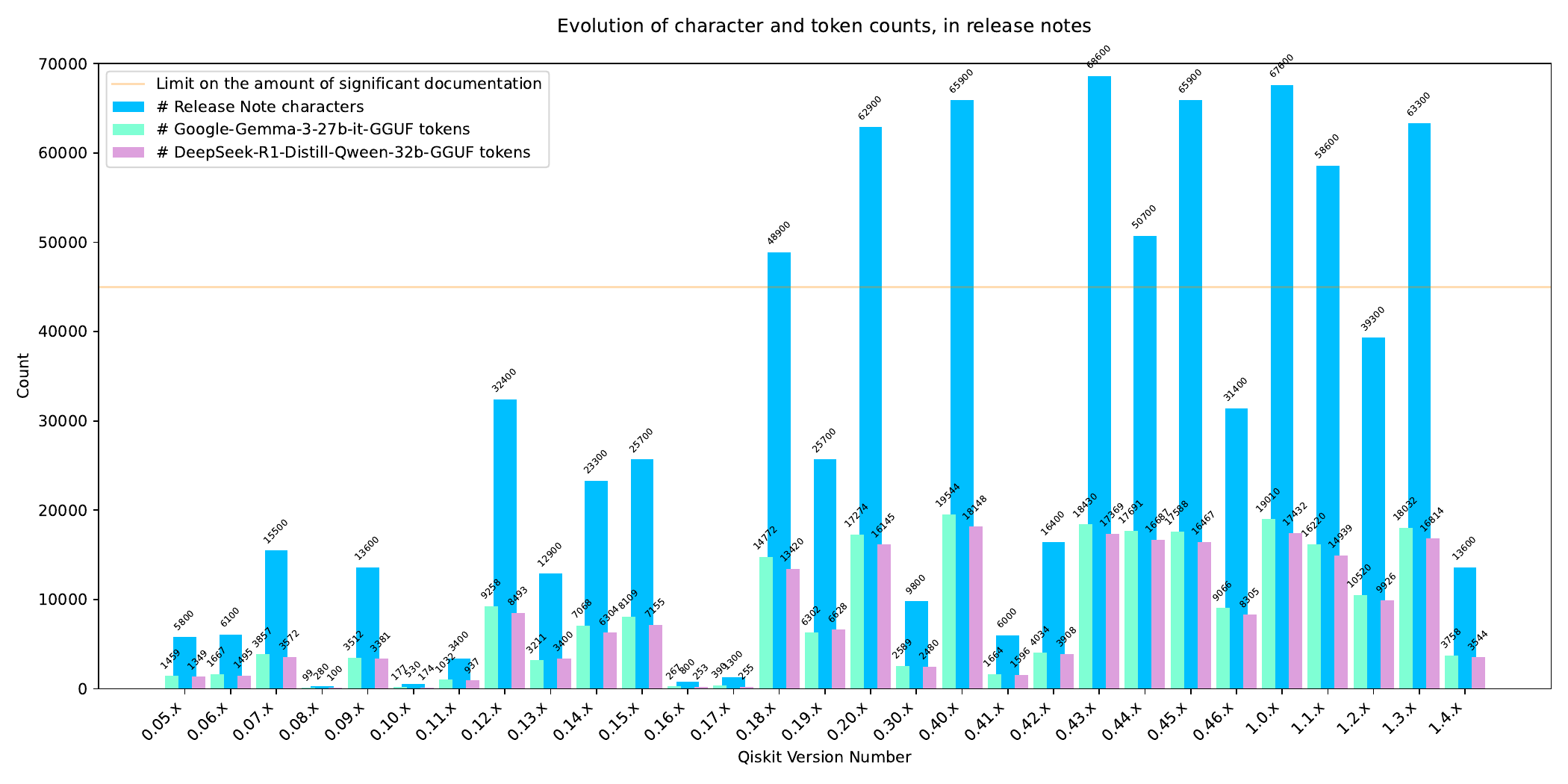}
    \caption{Qiskit release notes size distribution}
    \label{fig:qiskit_versions}
\end{figure}

However, careful prompting is needed in order to reduce ambiguities, replication, contradictions, or hallucinations, impacting the accuracy of the responses. Adding more contextual information directly into the prompt increase extensibility and completeness of the responses. In this regard, the need for complementary strategies to address constraints associated with context length of the current models also becomes evident. An analysis of RAG techniques that allows the models to access large amount of documentation is pending, but given the current context length of the models and the size in tokens of the documentation, it may be necessary to implement and asses. See Figure~\ref{fig:qiskit_versions}.

Another relevant topic is the knowledge of updates introduced by a specific Qiskit version to quantum components and associated performance. In this regard, the dimensions of our taxonomy “Migration Difficulty” and “Relation to QSE/SE” are intended to serve as the link for such types of assessments.

Having a complete taxonomy will allow us to measure the performance of LLMs, copilots and other automatic tools for code migration grouped by migration scenarios between Qiskit versions, and to prioritize the efforts towards the development of techniques and tools that solve the more difficult cases, or towards the QSE specific scenarios were no other tools are available.

We know LLMs have different performance depending on the programming language \cite{twist2025llmslovepythonstudy}. Python is one of the languages were most of LLMs excels at coding. Using LLMs to address migration scenarios in Qiskit could present an advantage over other quantum specific programming languages like Q\# which has C-like syntax. However, this claims needs validation. Our taxonomy can be the foundation for those experiments.

\section{Conclusion and Future Work}
\label{sec:futurework}

This work allowed us to identify the strengths of LLMs to generate a taxonomy of migration scenarios which was validated using an hybrid approach and comparing the output of the models with a manually generated taxonomy. The LLMs ability to detect cases not covered by an SE expert, in addition to greater descriptive precision and completeness in code examples that support each migration scenario, have proven useful enough to aid quantum software programmers to identify the performance gain, future deprecations, and improvement on circuit visualization easily. Given that these models can recognize the migration scenarios automatically, the next step is to asses the LLMs and copilots capabilities to solve this problems, also in an automatic or semi-automatic fashion.

The development of a taxonomy, automated or otherwise, allow us to move our research forward towards automatic tools for quantum software development. Out next steps are:
\begin{itemize}
\item Generate the taxonomy for all the Qiskit versions, including the newly released version 2.0.0. This step includes continue refining the prompts and testing models with larger context size, test more advanced techniques like prompt summarizing, Retrieval-Augmented Generation (RAG) \cite{lewis2021retrievalaugmentedgenerationknowledgeintensivenlp}, Chain of Thought (CoT)\\ \cite{wei2023chainofthoughtpromptingelicitsreasoning} and other reasoning architectures.  

\item  To evaluate the ability of LLMs to detect different migration scenarios in quantum software code and apply the migration automatically. We are interested in asses the LLMs performance regarding detecting and fixing the different migration categories, with special attention to those with high difficulty and those that are QSE specific. 

\item Survey or develop metrics that allow us to measure the performance of different models to detect and fix the migration scenarios. Ideally, a copilot may scan the code and detect migration scenarios and suggest ways to fix or improve. However, to check code accuracy in quantum software may require formal software verification techniques applied to QSE.

\end{itemize}

\bibliographystyle{apalike}
\bibliography{referencias2}

\end{document}